\documentclass[twocolumn,showpacs,preprintnumbers,amsmath,amssymb]{revtex4}

\usepackage{graphicx}
\usepackage{dcolumn}
\usepackage{bm}

\begin{document}

\preprint{APS/123-QED}


\title{Nonlinear  effects
in  microwave photoconductivity of
two-dimensional electron systems 
} 
\author{ 
V.~Ryzhii}
\email{v-ryzhii@u-aizu.ac.jp}
\affiliation{Computer Solid State Physics Laboratory, University of Aizu,
Aizu-Wakamatsu 965-8580, Japan}
\author{R.~Suris}
\affiliation{A.~F.~Ioffe Physical-Technical Institute,
Russian Academy of Sciences,  St.~Petersburg 194121, Russia 
}
\date{\today}

\begin{abstract}
We present a model for microwave photoconductivity of
two-dimensional electron systems in a magnetic field which describes
the effects of strong microwave and steady-state electric fields. 
Using this model, we derive
an analytical formula for 
the photoconductivity
associated with photon- and multi-photon-assisted impurity scattering
as a function of  the frequency and power of microwave radiation. 
According to the developed model,
the microwave conductivity is an oscillatory
function of the frequency of microwave radiation and the cyclotron
frequency which turns  zero  at the cyclotron
resonance and its harmonics. It exhibits maxima  and minima 
(with absolute negative conductivity)
at the microwave frequencies  somewhat different 
from the resonant frequencies.
The calculated power dependence of the amplitude
of the microwave photoconductivity oscillations 
exhibits pronounced sublinear behavior similar
to a logarithmic function. The height of the microwave
photoconductivity maxima and the depth of its minima
are nonmonotonic functions of the electric field.
It is pointed to the possibility of a strong widening
of the maxima and minima due to a strong sensitivity of their
parameters on the electric field and the presence of strong long-range
electric-field fluctuations.
The obtained dependences  are consistent with the results of the experimental
observations.
\end{abstract}



\maketitle

\section{Introduction}
Transport properties of two-dimensional electron systems (2DES's)
subjected to a transverse magnetic field 
were extensively studied in late 60s and early 70s.
To the best of our knowledge, the first paper on this topic was
published by Tavger and Erukhimov~\cite{1}.
In this paper, the dissipative conductivity (diagonal component 
of the conductivity tensor) associated with the tunneling
between the Landau levels (LL's) accompanied by
the impurity scattering
of electrons was calculated as a function of the electric and magnetic field.
 The  dissipative current-voltage characteristic
obtained considering the impurity scattering
in the Born approximation~\cite{1} 
is given by an non-analytical dependence $j \propto E^{-2}\exp(- E_c^2/2E^2)$,
where $E$ is the electric field,
$E_c = \hbar\Omega_c/eL$, $\hbar$ is the Planck constant, 
$\Omega_c$ and $L$ are the cyclotron frequency and quantum Larmor radius, 
respectively, and $e$ is the electron charge. Later, it was shown~\cite{2}
that more strict consideration of the impurity scattering leads to
the ``restoration'' of the Ohm's law, so that $j \propto E$ at 
$E < E_b = \hbar \Gamma/eL$, where $\hbar\Gamma$ is the LL-broadening, and 
$j \propto E^{-1}$ when $E_b < E \ll E_c$. 
Pokrovsky {\it et al.}~\cite{3} demonstrated that the inclusion of
the processes of the electron inter-LL tunneling via bound impurity
states can give rise to an exponential dependence similar to that obtained
in Ref.~\cite{1} but with a  smaller characteristic field $E_c$.
Theoretical studies
of the transport in a 2DES
irradiated with microwaves  or under non-equilibrium conditions
associated with intraband or intersubband optical excitation 
were carried out in Refs.~~\cite{4,5,6}. In particular, it was predicted
by Ryzhii~\cite{4}
(see also Ref.~\cite{7})
that microwave radiation with the frequency $\Omega$ somewhat
exceeding the cyclotron frequency or its harmonics $\Lambda\Omega_c$,
where $\Lambda$ is an integer, can result in the absolute
negative conductivity (ANC) in a 2DES. 
Nonlinear microwave photoconductivity associated with the effect of
photon-assisted impurity scattering of electrons 
was studied theoretically
in Refs.~\cite{8,9,10}. In particular, it was demonstrated
that the multi-photon processes (both virtual and real)
can significantly influence the dissipative and 
Hall components of the conductivity
tensor. 
Since current-voltage
characteristics with ANC  inevitably exhibit ranges with the negative differential
conductivity (NDC),
it was clear that time that uniform states of a 2DES with ANC can be 
unstable~\cite{11}.

An interest in theoretical studies of non-linear transport in 2DES's
has revived after the observation of the breakdown of 
the quantum Hall regime~\cite{12} (see, for example,  the articles by
Heinonen, {\it et al.}~\cite{13}, Balev~\cite{14}, 
Chaubet, {\it et al.}~\cite{15,16},
Komiyama, {\it et al.}~\cite{17,18}, and
review by Nachtwei~\cite{19}).

Recent observations of
vanishing electrical resistance/conductance in 2DES's
caused  by microwave radiation~\cite{20,21,22,23}, have stimulated 
extensive efforts to clarify the  nature
of the uncovered effects and triggered a new surge of theoretical
papers (see, for example,  Refs.~\cite{24,25,26,27,28,29,30,31,32,33}).
The occurrence of the so-called zero-resistance/conductance
states  is primarily considered~\cite{24,25,26,27,32} as a manifestation
of the effect of ANC associated with the photon-assisted impurity
scattering~\cite{4,7,26,32}, 
possibly, complicated by the scattering  processes involving
acoustic phonons~\cite{30,31,32,33}. 
The role of multi-photon-induced scattering processes was evaluated
in Refs.~\cite{28,29} using  models
similar to that considered earlier~\cite{8,9}. Vavilov and Aleiner~\cite{32} have developed a general
approach based on the quantum Boltzmann equation which provides
a description of non-linear effects in 2DES.   
In this paper, we use more simple and transparent model 
of the microwave conductivity
bearing in mind the goal to obtain explicit analytical formulas
describing non-linear effects, namely, 
formulas for the dependences of the photoconductivity maxima and minima 
on the microwave power   and the electric field (i.e., on the ac and dc
fields).
In particular, we demonstrate that the 
magnitude of the microwave photoconductivity maxima and minima is
a non-monotonic function of 
the microwave power   and the electric field  . 
It is also shown that an increase in
the microwave power   and/or the electric field leads to
a marked shift of the maxima and minima and an increase in
 their width. Due to a high sensitivity 
of the microwave photoconductivity to the local electric field,
the observable characteristics of the 2DES can be essentially
affected by long-range electric-field fluctuations.

The obtained results shed light on some features
of the effects observed experimentally.

After this introduction, in Sec.~II, we write down a general formula for
the dissipative current based on the notion that this current
is associated with the spatial displacement of the electron
Larmor orbit centers caused by the photon-induced impurity scattering
processes. The probability of these processes 
is a function of the net dc electric field 
(including both the applied and Hall components) 
and the ac microwave electric field.
This probability is presented as a sum
of the terms corresponding to the participation of different number
of real photons~\cite{8,9}. Such a technique allows 
to bypass the diagram (perturbation)
summation. 
In Sec.~III, we calculate the microwave photoconductivity
in relatively low dc electric field (Ohmic regime)
as a function of the microwave frequency
and power. Section~IV deals with the calculation
of the microwave photoconductivity in a strong dc electric field
when the latter substantially affects the scattering processes.
In Sec.~V, we consider the effect of microwave radiation on
intra-LL  impurity scattering.
Section~VI deals with the discussion of the obtained results
and its relevance to
the pattern of the experimental
observations.


\section{General equations}

At low temperatures $T \ll \hbar\Omega_c$ and low electric fields
$E \ll E_c$,
 the dissipative dark current (without irradiation)
is associated mainly with the electron transitions within the same  LL. 
Under the  microwave radiation,
the inter-LL electron photon-assisted
transition can markedly contribute to the
dissipative current. The microwave radiation can also affect
the intra-LL scattering processes.
Therefore, the microwave
photocurrent, i.e., the variation of the  dissipative
current cased by irradiation can be presented as

\begin{equation}\label{eq1}
j_{ph} = j_{ph}^{(inter)} +  j_{ph}^{(intra)}
,
\end{equation}
where $j_{ph}^{(inter)}$
is the contribution 
associated with the electron transitions between the LL's 
with different indices stimulated by microwave radiation, 
the second term in the right-hand side of Eq.~(1) is the variation 
of the intra-LL component
of the dissipative current.

For the  probability of the transition between
the $(N, k_x, k_y)$ and $(N^{\prime}, k_x + q_x, k_y + q_y)$ electron
states
in the presence of the net dc electric field ${\bf E} = (E,0,0)$
perpendicular to the magnetic field ${\bf H} = (0,0,H)$
and the ac microwave field ${\bf E}_{\Omega} = ({\cal E}e_x, {\cal E}e_y, 0)$
polarized in the 2DES plane ($e_x$ and $e_y$ are the components
of the microwave field polarization vector),
we will use the following formula
obtained on the base of the interaction representation
of the operator of current via solutions of
the classical equations of electron motion~\cite{8,9}:

$$
W_{N, k_x, k_y; N^{\prime}, k_x + q_x, k_y + q_y}
= \frac{2\pi}{\hbar}\sum_{M}
{\cal N}_i|V_q|^2|Q_{N,N^{\prime}}(L^2q^2/2)|^2
$$ 
\begin{equation}\label{eq2}
\times J_M^2(\xi_{\Omega}(q_x, q_y ))\,
\delta[M\hbar\Omega + (N - N^{\prime})\hbar\Omega_c + eEL^2q_y].
\end{equation}
Here $N$ is the LL index, $k_x$ and $k_y$ are the electron quantum numbers,
$(q_x$ and  $q_y)$ are their variations due to
photon-assisted impurity scattering, $q = \sqrt{q_x^2 + q_y^2}$, 
$e = |e|$ is the electron charge, 
${\cal N}_i$ is the impurity concentration, and $V_q \propto
q^{-1}\exp(- d_iq)$ is the matrix
element of the electron-impurity interaction, which accounts for
the localization of electrons in the $z$-direction, where $d_i$ is
the spacing between the 2DES and the $\delta$-doped layer.
The functions characterizing the overlap of the electron
initial and final states are
$Q_{N,N^{\prime}}(\eta) = P_N^{(N^{\prime} - N)}(\eta)exp(-\eta/2)$,
$P_N^{(N^{\prime} - N)}(\eta) \propto L_N^{(N^{\prime} - N)}(\eta)$,
where $L_N^{{\Lambda}}(\eta)$ is the Laguerre polynomial.
The LL form-factor is determined by the function
$\delta (\varepsilon)$, 
which at a small  broadening 
$\Gamma$ can be assumed to be the Dirac delta function.
The effect of microwave radiation is reduced to the inclusion
of the energy of really absorbed or emitted $M$ photons 
$M\hbar\Omega$ in the transition energy balance (in the argument 
of the function $\delta$ in Eq.~(2)) and  the appearance of the Bessel
functions $J_M(\xi_{\Omega}(q_x, q_y ))$, that reflexes the contribution
of virtually  absorbed and emitted photons~\cite{8,9}.
Here  
\begin{equation}\label{eq3}
\xi_{\Omega}(q_x, q_y) = \frac{e{\cal E}}{m}
\frac{|q_xe_x + q_ye_y - i(\Omega_c/\Omega)(q_xe_y - q_ye_x)|}{|\Omega_c^2 - \Omega^2|},
\end{equation}
where   $m$ is the electron effective mass.
Near the cyclotron resonance, the dependence  $\xi_{\Omega}(q_x, q_y)$
is close to isotropic. Disregarding the polarization
effects, we will assume  in the following that
\begin{equation}\label{eq4}
\xi_{\Omega}(q_x, q_y ) = \xi_{\Omega}\cdot Lq.
\end{equation}
Here
$\xi_{\Omega} = \sqrt{\langle |\xi_{\Omega}(q_x, q_y )|^2\rangle}/Lq$
, where the symbol $\langle...\rangle$
means averaging over the microwave field polarization,
so that $\xi_{\Omega} = (e{\cal E}\sqrt{\Omega_c^2 + \Omega^2}
/\sqrt{2}mL\Omega|\Omega_c^2 - \Omega^2|)$.
When  $\xi_{\Omega}$ becomes of order of unity,
the amplitude of the Larmor orbit center oscillation
in the microwave field is about  $L$.
Taking into account that for the transitions from
the LL's with $N \gg 1$, which play the main role in a 2DES
with a large filling factor, one can put
$P_N^{{\Lambda}}(\eta) \simeq J_{\Lambda}(2\sqrt{N\eta})$, 
the inter-LL contribution to the photocurrent can be presented
in the following form: 
\begin{figure*}[t]
\begin{center}
\includegraphics[width=15.cm]{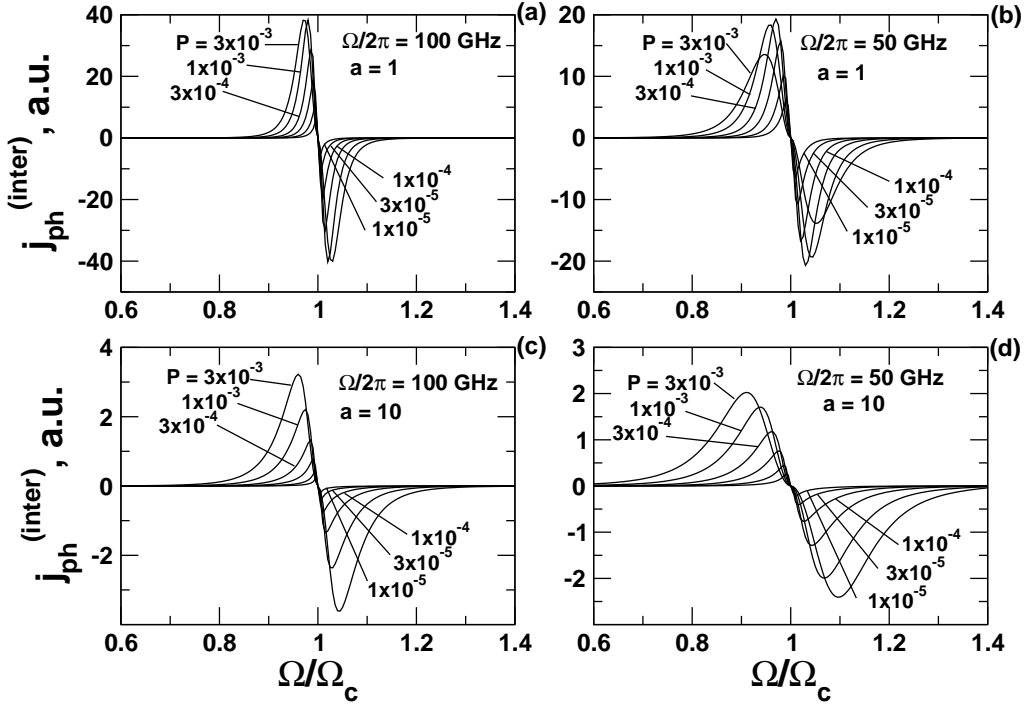}
\end{center}
\label{fig1}
\caption{
Dissipative photocurrent component
$j_{ph}^{(inter)}$ vs inverse cyclotron frequency $\Omega/\Omega_c$ 
in the vicinity
of the cyclotron resonance at different normalized
microwave powers $P$:
(a) $\Omega/2\pi = 100$~GHz,  $a = 1$,
(b)  $\Omega/2\pi = 50$~GHz, $a = 1$
(c) $\Omega/2\pi = 100$~GHz,  $a = 10$,  and  
(d) $\Omega/2\pi = 50$~GHz,   $a = 10$.
}
\end{figure*} 

$$
j_{ph}^{(inter)} \propto {\cal N}_iL^2\sum_{N,\Lambda, M \geq 0}
f_N(1 - f_{N + \Lambda})
$$
$$
\times\int dq_xdq_y \frac{q_y}{q^2} 
\exp\biggl(- 2d_iq - \frac{L^2q^2}{2}\biggr)
$$
\begin{equation}\label{eq5}
\times J_{\Lambda}^2(\sqrt{2N}Lq)J_M^2(\xi_{\Omega}Lq)) 
\delta(M\hbar\Omega - \Lambda\hbar\Omega_c + eEL^2q_y),
\end{equation}
where $f_N$ is the Fermi distribution function.

\section{Ohmic regime}

At $E < E_b$, assuming for definiteness
that $ \delta(\omega) = 
\Gamma/\pi(\omega^2 + \gamma^2)$, where $\gamma = \Gamma/\Omega_c$,
expanding the right-hand side of Eq.~(5) over
$eEL$, taking into account that  $J_{\Lambda}^2(\sqrt{2N}Lq) \simeq
\cos^2[\sqrt{2N}Lq - (2\Lambda + 1)\pi/4]/(\pi\sqrt{2N}Lq)$ for large $N$,
and integrating, we arrive at

\begin{equation}\label{eq6}
j_{ph}^{(inter)} \propto  E \Gamma
\sum_{\Lambda, M}
\frac{\Theta_{\Lambda}{\cal R}_M(\xi_{\Omega})(\Lambda\Omega_c - M\Omega)}
{[(\Lambda\Omega_c - M\Omega)^2 + \Gamma^2]^2}.
\end{equation}
Here 

\begin{equation}\label{eq7}
\Theta_{\Lambda} = \sum_{N}f_N(1 - f_{N + \Lambda})/\sqrt{N},
\end{equation}

\begin{equation}\label{eq8}
{\cal R}_M(z) = \int_{0}^{\infty}dx
J_M^2(z\,x)\exp(- \beta x -x^2/2)
, 
\end{equation}
where $\beta = 2d_i/L$. When $\beta < 1$, Eq.~(8) yields

\begin{equation}\label{eq9}
{\cal R}_M(z) \simeq 
\exp(- z^2)I_M(z^2),
\end{equation}
where $I_M(\eta)$ is the modified Bessel function.
In particular, at $z < 1$, function ${\cal R}_M(z)$
can be approximated as
\begin{equation}\label{eq10}
{\cal R}_M(z) 
\simeq \frac{\exp(- z^2)\,z^{2M}}{2^MM!}
\bigg[1 + \frac{z^{4}}{4(M+1)}\biggr].
\end{equation}
For $z \gg 1$, from Eq.~(9) one can obtain
\begin{equation}\label{eq11}
{\cal R}_M(z) \simeq \frac{1}{\sqrt{2\pi}z}.
\end{equation}

Taking into account that ${\cal E}^2 \propto p_{\Omega}$, 
where $p_{\Omega}$ is the microwave power density,  
introducing the characteristic microwave power 
${\overline p_{\Omega}} = 
m\Omega^3/
2\pi \alpha
\simeq 21.8m\Omega^3$,
where $\alpha = e^2/\hbar c \simeq 1/137$ and
$c$ is the speed of light, 
and using 
Eqs.~(6) and ~(9), for a 2DES with  $\beta < 1$  we arrive at

$$
j_{ph}^{(inter)} \propto  E\Gamma
\exp\biggl[- P\,f\biggl(\frac{\Omega}{\Omega_c}\biggr)\biggr]
$$
$$
\times \biggl\{I_1\biggl(P\,f\biggl(\frac{\Omega}{\Omega_c}\biggr)\biggr)
\sum_{\Lambda}
\frac{\Theta_{\Lambda}(\Lambda\Omega_c - \Omega)}
{[(\Lambda\Omega_c - \Omega)^2 + \Gamma^2]^2},
$$
\begin{equation}\label{eq12}
 + I_2\biggl(P\,f\biggl(\frac{\Omega}{\Omega_c}\biggr)\biggr)
\sum_{\Lambda}
\frac{\Theta_{\Lambda}(\Lambda\Omega_c - 2\Omega)}
{[(\Lambda\Omega_c - 2\Omega)^2 + \Gamma^2]^2} +...\biggr\},
\end{equation}
where $P = p_{\Omega}/{\overline p_{\Omega}}$ is the normalized microwave
power
and
$f(\omega) = \omega(1 + \omega^2)/(1 - \omega^2)^2$.
In particular, at $P\,f(\Omega/\Omega_c) < 1$, Eq.~(12)
can be presented as

$$
j_{ph}^{(inter)} \propto  E\Gamma
\exp[- P\,f(\Omega/\Omega_c)]
$$
$$
\times \biggl\{\frac{P}{2}\,f\biggl(\frac{\Omega}{\Omega_c}\biggr)
\sum_{\Lambda}
\frac{\Theta_{\Lambda}(\Lambda\Omega_c - \Omega)}
{[(\Lambda\Omega_c - \Omega)^2 + \Gamma^2]^2},
$$
\begin{equation}\label{eq13}
 + \frac{P^2}{8}\,f^2\biggl(\frac{\Omega}{\Omega_c}\biggr)
\sum_{\Lambda}
\frac{\Theta_{\Lambda}(\Lambda\Omega_c - 2\Omega)}
{[(\Lambda\Omega_c - 2\Omega)^2 + \Gamma^2]^2} +...\biggr\}.
\end{equation}

One can see from the first term in the right-hand side of
Eqs.~(12) and (13) that $j_{ph}^{(inter)} = 0$
at $\Lambda\Omega_c = \Omega$, 
 whereas
$j_{ph}^{(inter)} > 0$ if
$\Lambda\Omega_c$ slightly  exceeds $\Omega$  
and  $j_{ph}^{(inter)} < 0$ (i.e., the dissipative 
microwave photocurrent is in opposition to the electric field) when
$\Lambda\Omega_c$ is somewhat smaller than $\Omega$.
 As it also follows from Eqs.~(12) and (13), 
$ j_{ph}^{(inter)}$ exhibits  maxima and minima
at $\Omega/\Omega_c = \Lambda - \delta^{(+)}$ and 
$\Omega/\Omega_c = \Lambda + \delta^{(-)}$, respectively, where 
$0 < \delta^{(+)},\,\delta^{(-)} < 1$.
These maxima and minima correspond to the electron transitions
with the absorption of one real photon.
Apart from the single-photon maxima and minima,
there are the maxima and minima associated with the two-photon
 (described by the second terms in the right-hand sides
of Eqs.~(12) and (13)) and multiple-photon absorption processes.

Using Eqs.~(12) or (13), one can find that
$\delta^{(\pm)} \simeq \sqrt{3}\Gamma/\hbar\Omega_c$. 
At moderate microwave powers,
estimating the LL broadening as (see, for example, Refs.~\cite{29,34,35}) 
$\Gamma = \sqrt{2\Omega_c a/\pi\tau}$,
where $\tau$ is the electron momentum relaxation 
time estimated from the electron mobility at $H = 0$ and $p_{\Omega} = 0$
and $a \geq 1$ is a semi-empirical broadening
parameter which describes
the difference between the scattering time $\tau$ and the net scattering
time determined by all other mechanisms,
one can find that $\delta^{(+)} \simeq \delta^{(-)}
\simeq \sqrt{6a/\pi\Omega_c\tau}$.
Assuming $\tau = 5.8\times10^{-10}$~s, $a = 1 - 10$,
 and $\Omega/2\pi = 50$~GHz (as in
Ref.~\cite{23}) and using the above analytical estimate,
we obtain 
$\delta^{(+)} \simeq \delta^{(-)} \simeq 0.1 - 0.3$.

Figures~1 and 2 show the dependences 
of the inter-LL component of the microwave photocurrent
$j_{ph}^{(inter)}$ on the inverse cyclotron frequency $\Omega/\Omega_c$
for different broadening parameters and
for different  microwave powers and  frequencies
calculated for a 2DES with $\beta < 1$
using Eq.~(12). 
It is assumed that  $\tau = 5.8\times10^{-10}$~s.
The oscillatory $j_{ph}^{(inter)}$ vs $\Omega/\Omega_c$ dependence
at different $P$ with  the paired maxima and minima
corresponding to the cyclotron resonance and its harmonics 
is shown in Fig.~2. One can see from Figs.~1 and 2 that the 
height of  the maximum  and the depth of the minimum near 
the cyclotron resonance (as well as near the cyclotron harmonics)
increase with increasing microwave power. This increase is rather slow; 
the span of the photoconductivity as a function of
the microwave power at low and moderate powers
behaves  similar to a logarithmic
function. However, at large powers, height of  the maximum  
(depth of the minimum) saturates and begins to fall (see Fig.~3).
Figure~3 demonstrates the variations
of the photoconductivity
maxima and minima  with increasing microwave
power. As seen from Fig.~3, the height of the first one-photon
maximum (depth of the relevant minimum), corresponding
to $\Lambda = 1$ and $M = 1$, falls when
the microwave power increases beyond some threshold value.
In this range of microwave power, the maxima (minima)
corresponding to higher resonances can be comparable with that
near the cyclotron resonance. 
At high microwave powers, the two-photon resonant maxima (minima)
can increase faster than those associated with the one-photon
absorption processes (compare the curve for 
$\Omega/\Omega_c \pm \delta^{(\pm)} = 1.5$ and the curves
for $\Omega/\Omega_c \pm \delta^{(\pm)} = 1$ and 2).
In particular, 
when   approximation~(9) is valid,
the ratio of  the two-photon maxima 
$j_{ph}^{(2,3)}$ ( $M = 2$ and $\Lambda = 3$)
to the single-photon maxima $j_{ph}^{(1,2)}$
($M = 1$ and $\Lambda = 2$)
calculated
using Eq.~(12) can be estimated as
${\rm max}\,j_{ph}^{(2,3)}/{\rm max}\,j_{ph}^{(1,2)} \simeq 
3.37P e^{2P}$.
As it follows from the latter estimate, the  ratio in question,
 being rather small at 
small $P$,  markedly rises with increasing $P$.
Figure~4 shows the variations of the position of the photoconductivity
maxima near the cyclotron resonance $\delta^{(+)}$
and its width at half-maximum $\Delta^{(+)}$ 
(normalized by $\Omega_c$) with increasing normalized
microwave power $P$.
One can also see from Figs.~1, 2, and 4 that the width of the resonant maximum
and minimum  increases with increasing microwave power.
The broadening of the cyclotron absorption line at heightened
microwave power associated with similar mechanism
was discussed recently~\cite{35}.
Apart from this, at high microwave powers, 
the transition rate   and, therefore,
the LL broadening become larger. 
This can lead to an increase in the broadening parameter
$a$ and, hence, in an extra increase
in the maximum and minimum width 
when the microwave power increases.
At rather high powers, the  two-photon resonances can be observable.
Figure~2 exhibits comparably weak maximum and minimum 
in the vicinity of  $\Omega/\Omega_c = 1.5$ corresponding to
the two-photon transition ($M = 2$ and $\Lambda = 3$).
These results are consistent
with experimental data by Mani, {\it et al.}~\cite{20,23} 
and Zudov, {\it et al.}~\cite{21}
(see also Ref.~\cite{36}). 
Assuming, as in Ref.~\cite{21}, that the power of the  
microwave source and the sample cross-section are 
10 - 20~mW and  0.25~cm~$^{-2}$, respectively,
and using the above formula for the characteristic microwave power 
${\overline p_{\Omega}} $,
for $\Omega/2\pi = 50$~GHz one can find that the experimental conditions
correspond to max~$P \lesssim (1 - 2)\times10^{-2}$. The 
later values are on the order of or larger than those used in Figs.~1 and 2.

\begin{figure}[t]
\begin{center}
\includegraphics[width=8.7cm]{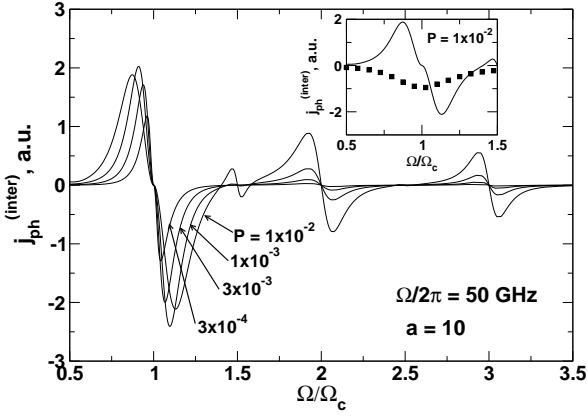}
\end{center}
\label{fig2}
\caption{Oscillations of 
$j_{ph}^{(inter)}$ as a function of inverse cyclotron frequency
$\Omega/\Omega_c$ 
at different microwave powers ($\Omega/2\pi = 50$~GHz, $a = 10$).
Inset shows $j_{ph}^{(inter)}$ (solid line) and  $j_{ph}^{(intra)}$ 
(squares) vs
inverse cyclotron frequency. 
}
\end{figure} 

As follows from Eqs.~(6) and (8), the height of the maxima and the
depth  of the minima determined by the function ${\cal R}_M(z)$
depend on the parameter $\beta$, i.e., on the thickness of the spacer
separating the 2DES and the donor layer. This function is 
plotted in Fig.~5 for $M = 1$ (one-photon absorption) and $M = 2$
(two-photon absorption). One can see that  ${\cal R}_1(z)$
and  ${\cal R}_2(z)$ pronouncedly decrease with increasing $\beta$.
If  $\beta \gg 1$, using Eqs.~(6) and (8),
one can obtain instead of Eq.~(12)

\begin{figure}[t]
\begin{center}
\includegraphics[width=8.1cm]{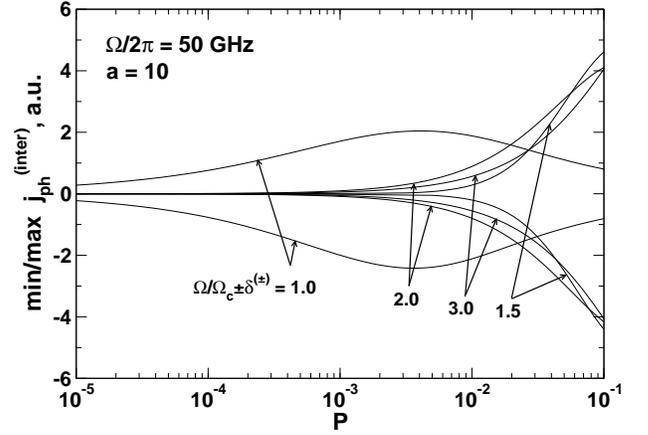}
\end{center}
\label{fig3}
\caption{
Maxima and minima of
$j_{ph}^{(inter)}$ corresponding to
different resonances as functions of normalized microwave power~$P$
($\Omega/2\pi = 50$~GHz and $a = 10$).
}
\end{figure}
\begin{figure}[b]
\begin{center}
\includegraphics[width=8.3cm]{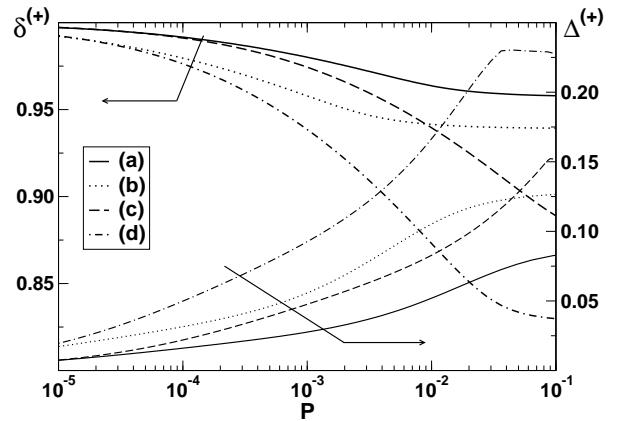}
\end{center}
\label{fig4}
\caption{Position $\delta^{(+)}$ and 
width at half-maximum $\Delta^{(+)}$
of near cyclotron maximum vs normalized microwave power $P$. Curves (a) - (d)
correspond to photocurrent spectral dependences in Figs.~1(a) - 1(d).}
\end{figure}
$$
j_{ph}^{(inter)} \propto  \frac{E\Gamma}{\beta}
\biggl\{{\cal F}_1\biggl(\frac{\sqrt{P}}{\beta}
\sqrt{f\biggl(\frac{\Omega}{\Omega_c}\biggr)}\biggr)
\sum_{\Lambda}
\frac{\Theta_{\Lambda}(\Lambda\Omega_c - \Omega)}
{[(\Lambda\Omega_c - \Omega)^2 + \Gamma^2]^2}
$$
\begin{equation}\label{eq14}
 + {\cal F}_2\biggl(\frac{\sqrt{P}}{\beta}\sqrt{f\biggl(\frac{\Omega}{\Omega_c}\biggr)}\biggr)
\sum_{\Lambda}
\frac{\Theta_{\Lambda}(\Lambda\Omega_c - 2\Omega)}
{[(\Lambda\Omega_c - 2\Omega)^2 + \Gamma^2]^2} +...\biggr\}
\end{equation}

Here for $\beta \gg 1$
\begin{equation}\label{eq15}
{\cal F}_M(z) = \int_0^{\infty}dx J_M^2(z\,x)\exp(-x) \simeq \beta 
{\cal R}_M(z/\beta).
\end{equation}
At not too large 
$z$,
 setting
$J_M(z) \simeq a_{M}
\sin(\pi z\,x/b_{M})$, where 
$a_{M}$ is the first maximum of the pertinent Bessel function
and $b_{M}$ corresponds to its first zero, one can obtain
\begin{equation}\label{eq16}
{\cal F}_M(z) \simeq 
\biggl(\frac{\pi a_M}{2b_M}\biggr)^2\frac{z^2}
{[1 + (\pi/b_M)^2z^2]}.
\end{equation}
In particular, for $M = 1$ Eq.~(16) yields
$F_M(z) \propto  z^2/\beta(1 + 0.67z^2)$. At very large $z$,
$F_M(z)$ becomes a decreasing function of $z$
similar to that given by Eq.~(11).
Considering Eq.~(16), near
the first one-photon resonance
($\Lambda = 1$ and  $M = 1$) at $\beta \gg 1$, from Eq.~(14) we obtain

$$
j_{ph}^{(inter)} \propto 
\frac{ E\Gamma P}{\beta^3}\,f\biggl(\frac{\Omega}{\Omega_c}\biggr)
\biggl[1 + 0.67\frac{P}{\beta^2}\,f\biggl(\frac{\Omega}{\Omega_c}\biggr)\biggr]^{-1}
$$
\begin{equation}\label{eq17}
\times\frac{(\Omega_c - \Omega)}
{[(\Omega_c - \Omega)^2 + \Gamma^2]^2}.
\end{equation}
Equation~(14) demonstrates a decrease
in  the microwave
photoconductivity and the height of its maxima (depth of the minima)
with increasing parameter $\beta$ and
slowing down 
with increasing power. The former is due to a decrease in
the impurity scattering rate because of a smoothening
of the fluctuating electric field created by
charged impurities when the spacer becomes thicker
($d_i$ becomes larger). One can see a similarity
between the roles of the inverse parameter $\beta^2$ and
the normalized microwave power $P$. 

\begin{figure}[t]
\begin{center}
\includegraphics[width=7.5cm]{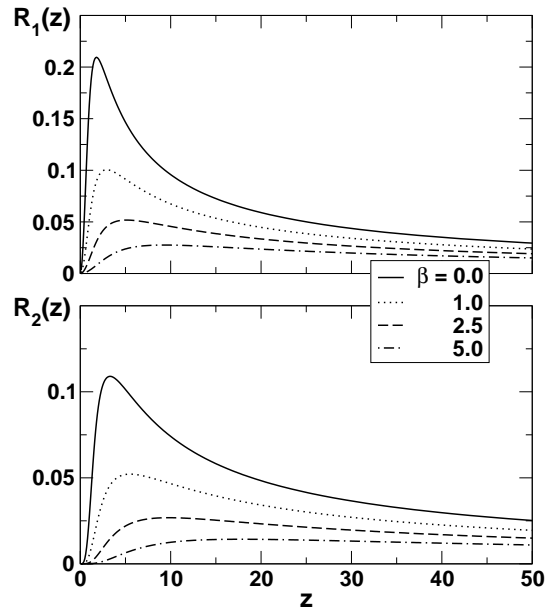}
\end{center}
\label{fig5}
\caption{Functions ${\cal R}_1(z)$ and ${\cal R}_2(z)$ for different
values of parameter $\beta$.}
\end{figure}

\section{Photoconductivity in strong electric field}

The average electric field $\overline {E}$ in the experimental 
situations~\cite{20,21,22,23}
is rather moderate, so one can assume that $\overline {E} < E_b$.
However, under the condition of ANC, electric-field domain
structures can be formed~\cite{25,27,33}. The value of the
electric field $E_0$, at which
the net dissipative conductivity changes its sign from negative
to positive, essentially affects the magnitude of the electric
field variations in the domain structures in question.
Although it is unclear yet what
mechanism determines  the threshold electric field $E_0$,
one can assume that it can be much larger than $E_b$.
In this case, 
the  electric field in some regions may significantly
 exceed $E_b$. 
As shown by Shchamkhalova
and the authors~\cite{37}, the LL broadening $\hbar\Gamma$ and, therefore,
the characteristic field $E_b$ 
can substantially decrease
in the electric field. 
Moreover, recent experiments~\cite{38}
 showed that strong long range ($\lambda \gg L$)
 fluctuations are present in 2DES's with high electron mobility.
Due to these fluctuations, the local electric field can
be of  a rather large magnitude (about 30 - 150~kV/cm, 
as estimated by Kawano, {\it et al.}~\cite{18}), tangibly affecting
the electron inter-LL transitions~\cite{1,4,13,17}. 
Hence, the inequalities $E > E_b$ or $E \gg E_b$ can take place
in the 2DES under consideration.
\begin{figure}[t]
\begin{center}
\includegraphics[width=8.0cm]{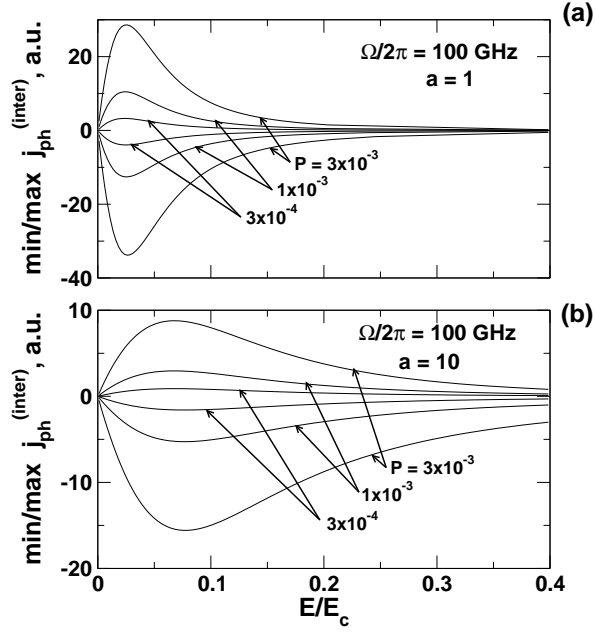}
\end{center}
\label{fig6}
\caption{
Near cyclotron maxima and minima of
$j_{ph}^{(inter)}$ vs electric field at
different  microwave powers for
$\Omega/2\pi = 100$~GHz and different broadening parameter
$a$: (a) $a = 1$ and (b)~$a =~10$. 
}
\end{figure}

At relatively large net dc electric fields $E > E_b$,
$\delta (\omega)$ can be considered
as the Dirac $\delta$-function. Taking into account that the
transitions between
high LL's
provide main contribution to the mechanism of
microwave photoconductivity under consideration, and 
integrating over $q_y$ in Eq.~(5), we obtain

$$
j_{ph}^{(inter)} \propto 
\sum_{\Lambda, M }
\Theta_{\Lambda} {\cal R}_M\biggl(\xi_{\Omega},
\frac{\hbar|\Lambda\Omega_c - M\Omega|}{eEL}\biggr)
$$
\begin{equation}\label{eq18}
\times\biggl[\frac{\hbar(\Lambda\Omega_c - M\Omega)}{eE^2L}\biggr]
\exp\biggl[- \frac{\hbar^2(\Lambda\Omega_c - M\Omega)^2}{2(eEL)^2}\biggr],
\end{equation}
where
$$
{\cal R}_M(z,y) = \int_0^{\infty} dx J_M^2(z\sqrt{x^2 + y^2})
$$
\begin{equation}\label{eq19}
\times\frac{\exp[- \beta\sqrt{x^2 + y^2}- 
(x^2 + y^2)/2]}
{(x^2 + y^2)^{3/2}}.
\end{equation}
At small $\xi_{\Omega}$,
Eq.~(18)  coincides with that obtained a long time ago~\cite{4}.
As follows from Eq.~(18), the microwave conductivity in strong electric field
also exhibits an oscillatory behavior with maxima and minima
at $\Omega/\Omega_c = \Lambda - \delta^{(+)}$ and 
$\Omega/\Omega_c = \Lambda + \delta^{(-)}$, respectively,
for the one-photon transitions , and 
$\Omega/\Omega_c = [\Lambda - \delta^{(+)}]/M$ and 
at $\Omega/\Omega_c = [\Lambda + \delta^{(-)}]/M$ for the
transitions with absorption of $M$ photons. Here
$\delta^{(+)} \simeq \delta^{(-)} \simeq eEL/\hbar\Omega_c = E/E_c$.
The width of the maxima and minima in question linearly increases with $E$
as $\Delta^{(+)} \simeq \Delta^{(-)} \simeq 2.24  E/E_c$,
while
their height is proportional to $E^{-1}$.
Indeed, Eq.~(18) yields the following dependence of 
the min/max~ $j_{ph}^{inter}$ for the one-photon
absorption near the cyclotron resonance on the microwave power $P$:
\begin{equation}\label{eq20}
{\rm min/max}\, j_{ph}^{inter} \propto 
\frac{\exp(- P\,f^{(\mp)})}{E}
I_1(P\,f^{(\mp)}),
\end{equation}
where $f^{(\pm)} = f(1 \pm E/E_c)$.
Here, assuming $\beta < 1$, we have used the following estimate: 
${\cal R}_1(z, 1) \simeq 0.4\,\exp(-z^2)I_1(z^2)$.
Figure~6 shows the dependences of min/max~$j_{ph}^{inter}$ on the electric
field at different microwave power calculated using an interpolation 
formula leading to  min/max~$j_{ph}^{inter}$ following from
Eq.~(12) in the
low-field region (for $\Omega/2\pi = 100$~GHz and $a = 1$ and $a = 10$)
and to Eq.~(20) in the high-field region.

A marked sensitivity of the resonant maxima and minima to the electric field
can be  crucial
(together with the LL-broadening associated with the interactions other
than the impurity scattering) for the explanation of their fairly wide width
observed experimentally, particularly, considering
strong electric-field fluctuations~\cite{38}.

\section{Suppression of Intra-LL photoconductivity}

Taking into account  that, as follows from  Eq.~(2),
the probability of the elastic impurity
scattering in the presence of  microwave radiation
differs from that in its absence  by
the factor $J_0^2(\xi_{\Omega}(q_x, q_y ))$ 
and generalizing the results of Ref.~\cite{2} (see also Ref.~\cite{35}),
the dissipative current associated with the intra-LL impurity
scattering can  given by

$$
j^{(intra)} \propto E  {\cal N}_iL^2\sum_{N}
f_N(1 - f_{N})
$$
$$
\times\int dq_xdq_y \frac{q_y^2}{q^2} 
\exp\biggl(- 2d_iq - \frac{L^2q^2}{2}\biggr)
$$
\begin{equation}\label{eq21}
\times J_{0}^2(\sqrt{2N}Lq)J_0^2(\xi_{\Omega}Lq)) 
\end{equation}
at $E < E_b$, and
$$
j^{(intra)} \propto \frac{{\cal N}_i}{E} \sum_{N}
f_N(1 - f_{N})
$$
$$
\times\int dq_xdq_y \frac{1}{q^2} 
\exp\biggl(- 2d_iq - \frac{L^2q^2}{2}\biggr)
$$
\begin{equation}\label{eq22}
\times J_{0}^2(\sqrt{2N}Lq)J_0^2(\xi_{\Omega}Lq)) 
\end{equation}
at  $E \gg E_b$.
Using the same procedure as in Sec.~III,
Eq.~(21) for the case $\beta < 1$ and
$E < E_b$,  can be reduced to
\begin{equation}\label{eq23}
j^{(intra)} \simeq j_{dark}^{(intra)}{\cal R}_0(\xi_{\Omega})
$$
$$
\simeq  j_{dark}^{(intra)}
\exp\biggl[- P\,f\biggl(\frac{\Omega}{\Omega_c}\biggr)\biggr]\,
\cdot I_0\biggl(P\,f\biggl(\frac{\Omega}{\Omega_c}\biggr)\biggr),
\end{equation}
where $j_{dark}^{(intra)}= j^{(intra)}|_{P = 0}$ is the dissipative
current associated with the intra-LL impurity scattering 
without irradiation (dark current).
Using Eq.~(23), we arrive at the following expression  for the
photoconductivity associated with the microwave-stimulated
variation of
the dissipative intra-LL conductivity:
$$
j_{ph}^{(intra)} \simeq
j_{dark}^{(intra)} \biggl\{
\exp\biggl[- P\,f\biggl(\frac{\Omega}{\Omega_c}\biggr)\biggr]
\cdot I_0\biggl(P\,f\biggl(\frac{\Omega}{\Omega_c}\biggr)\biggr) - 1\biggr\}
$$
\begin{equation}\label{eq24}
\simeq - \frac{3 }{4}j_{dark}^{(intra)}P 
\,f\biggl(\frac{\Omega}{\Omega_c}\biggr).
\end{equation}
The last term in the right-hand side of Eq.~(24) 
is valid when $P\,f(\Omega/\Omega_c) < 1$. 
As seen from Eq.~(24), the contribution
of the effect of microwave radiation on the elastic
intra-LL impurity scattering to the photoconductivity
is negative.  
This contribution as a function of $\Omega/\Omega_c$
exhibits a smeared minimum at the cyclotron
resonance. 
This minimum is attributed to the suppression of the impurity scattering
by microwave radiation. Such a suppression (associated
with the absorption and emission of virtual phonons) is most effective
when the amplitude of the Larmor orbit center
oscillation is maximum, i.e., at the cyclotron resonance.
The $j_{ph}^{(intra)}$  vs  
$\Omega/\Omega_c$ dependence calculated using Eq.~(24) is shown
by squares in 
the inset in Fig.~2.
Since this dependence does not comprises any resonant factor,
it can not affect significantly the oscillatory dependences
associated with the photon-assisted inter-LL transitions,
although it leads to some deepening of the photoconductivity minimum
near the cyclotron resonance.
Formulas~(23) and (24) correspond to the total suppression of
the dissipative intra-LL conductivity, i.e. to $j^{(intra)} = 0$ 
and therefore, $j_{ph}^{(intra)} = -
j_{dark}^{(intra)}$,  at the cyclotron resonance.
However, the dephasing of the electron oscillatory movement
in the microwave field due to different scattering events
can limit the suppression of the
dissipative intra-LL conductivity.
This effect can be included by the substitution of the function 
$f^{*}(\omega) = f(\omega)(1 - \omega^2)/
[(1 - \omega)^2 + \gamma^2]$
for the  function $f(\omega)$ introduced in Sec.~III. 
Taking into account that $f^*(1) = 1/2\gamma^2$,
at $P < 2(\Gamma/\Omega_c)^2$ we obtain 
min~$j^{(intra)} \simeq j_{dark}^{(intra)}[1 - P(\Omega_c/\sqrt{2}\Gamma)^2]$  and,  consequently,
min~$j_{ph}^{(intra)} \simeq 
- j_{dark}^{(intra)} P(\Omega_c/\sqrt{2}\Gamma)^2$.
At $P \gg 2(\Gamma/\Omega_c)^2$, one obtains
min~$j^{(intra)} \simeq j_{dark}^{(intra)}(\Gamma/\sqrt{\pi P}\Omega_c)$ 
and 
min~$j_{ph}^{(intra)} \simeq - j_{dark}^{(intra)}[1 - (\Gamma/\sqrt{\pi P}\Omega_c)] \simeq - j_{dark}^{(intra)}$.

\section{Comments}

The proposed model describes the following features
of the microwave conductivity in a 2DES subjected to a magnetic field:

(i) The zeroth microwave conductivity at the cyclotron
resonance and its harmonics, as well as
 the  location of the photoconductivity 
maxima and minima
in the vicinity of the resonances with ANC in the minima.

(ii) Nonlinear dependencies of the photoconductivity on the microwave power
characterized by slowing down,  saturation, and even decrease
in the minima/maxima magnitude with increasing  microwave power.

(iii) Shift of the photoconductivity maxima and minima and their
broadening with increasing microwave power and electric field.

(iv) Possibility of the  suppression of the dissipative
conductivity associated with the intra-LL transitions by the microwave 
radiation.

The positions of the photoconductivity
zeros, maxima, and minima
 are determined  by   the specific features of the photon-assisted
impurity scattering of electron in the magnetic field.
Some of these features of the microwave photoconductivity, 
predicted theoretically~\cite{4,7} and observed experimentally~\cite{21,23}
(see also Refs.~\cite{21,22,36}),
have been discussed in the framework of different 
theoretical models~\cite{24,26}.
It is instructive that the effect 
of suppression of the impurity intra-LL scattering
by microwave radiation can result in a  shift of the microwave
photoconductivity zero to  $\Omega/\Omega_c$ slightly smaller than unity. 

The photon-assisted scattering on acoustic phonons
also can lead to the oscillatory dependence of the dissipative conductivity
with ANC at $\Omega/\Omega_c$ between the cyclotron resonances and
its harmonics~\cite{30,31}. It is remarkable that this mechanism
provides the photoconductivity minima and maxima approximately
at the same values of  $\Omega/\Omega_c$,  where the photon-assisted impurity scattering yields, in contrast, maxima and minima.
Thus, the photon-assisted acoustic scattering can to some extent
suppress the oscillations associated with the photon-assisted
impurity scattering. Such a competition of the mechanisms
in question can be essential because
the piezoelectric  acoustic scattering is one of the main scattering
mechanism limiting the electron mobility in perfect 2DES's at low
temperatures~\cite{39}. The contribution of the photon-assisted
acoustic scattering mechanism to the microwave conductivity markedly
increases with the temperature, particularly, in the range
where $T$ is comparable with the characteristic energy of acoustic phonons
$\hbar s/L$ ($s$ is the speed of sound). Assuming $s = 3\times10^5$
and $H = 1 - 2$~kG, one can obtain $T_{ac} \simeq 0.4$~K. The latter
value is only slightly smaller than $T$ in the experiments. 
Hence, the suppression of the microwave photoconductivity
oscillations and the effect of ANC (a pronounced decrease 
in the minima/maxima magnitude) with increasing temperature
observed experimentally~\cite{20,21} can possibly  be attributed to
the inclusion of the abovementioned photon-assisted
acoustic scattering mechanism.
 
The photoconductivity considered 
above (and in Refs.~\cite{4,7,26,28,29,32})
is
associated with a direct effect of the microwave ac field on the
the  impurity  scattering of electrons. However, the photon-assisted
impurity scattering is accompanied by the absorption of photons
and, hence, some heating of the 2DES. Such a heating can lead to a
variation of the dissipative conductivity associated with
the scattering on impurities and acoustic phonons.
The dissipative current associated with the intra-LL impurity
scattering is sensitive 
to the temperature (mainly via the factor $\Theta_0$).
This effect in part gives rise to  the smearing of the Shubnikov-de Haas
oscillation with increasing temperature. However, the temperature
dependence of $\Theta_0$ in the range where the oscillations of microwave
photoconductivity under consideration were observed experimentally
is rather weak. The electron heating caused by the photon-assisted
impurity scattering processes is compensated by
the relaxation processes with the electron transitions from
higher to lower LL's with the emission of acoustic phonons.
As shown recently~\cite{40}, these relaxation processes contribute to
the dissipative conductivity (their contribution is negative)
complicating the spectral
dependence of the microwave photoconductivity.

A nontrivial dependences of the maxima/minima span 
on the microwave power are consistent with
those observed experimentally~\cite{20,23} and 
are attributed to the multi-photon
effects including the absorption and emission of real and virtual
phonons (see Refs.~\cite{8,9,10,29,35}).
A relatively large width of the microwave photoconductivity
maxima and minima is often identified with an extra 
broadening  of LL's (in comparison with
the LL broadening
corresponding to the electron
collision time determining the electron mobility)
associated with more complex interaction discussed previously
~\cite{24,26,29}.
As pointed out  above, 
this  effect can be attributed to
the nonlinear effects of the ac microwave field and
strong electric-field dependences of the local photoconductivity
characteristics combined with strong long-range fluctuations of
the electric field. In the presence
of strong 
long range ($\lambda \gg L$) electric-field
fluctuations the macroscopic properties of the 2DES, in particular,
the domain structures with the scale exceeding $\lambda$ can
be determined by the  components of the dissipative
photocurrent  averaged over these fluctuations
$\overline{j_{ph}(E)E_x/E}$
and $\overline{j_{ph}(E)E_y/E}$. Due to a strong electric
field dependence of $j_{ph}$ (see Eq.~(18)), the dependences of the 
averaged quantities on the components of the
averaged electric field  $\overline{E_x} \propto V$ and 
$\overline{E_y} \propto V_H$ (where $V$ and $V_H$ are the potential
drop along the current and the Hall current, respectively)
as well as their spectral dependences can be significantly
different from those of the local dissipative current.

\section*{Acknowledgments}
The authors are grateful to A.~Satou for essential assistance,
V.~Volkov and V.V'yurkov for useful discussions, and B.~Shchamkhalova
for providing materials related to Ref.~\cite{10}.

\end{document}